\documentclass[showpacs,pre,aps,twocolumn,superscriptaddress]{revtex4-1}

\pdfoutput=1

\usepackage{amsmath,amsfonts,amssymb,bm,graphicx,hyperref}
\usepackage{color}
\usepackage{array}

\begin{document}

\newcommand{\LL}{\mathbb{L}}
\newcommand{\HH}{\mathbb{H}}
\newcommand{\TT}{\mathcal{T}}
\newcommand{\ee}{\mathrm{e}}

\newcommand{\beq}{\begin{equation}}
\newcommand{\eeq}{\end{equation}}

\newcommand{\jj}{A}

\newcommand{\ddiv}{\operatorname{div}}

\newcommand{\rhobar}{\overline{\rho}}

\newcommand{\CC}{\mathcal{C}}
\newcommand{\Ac}{\mathcal{A}}

\newcommand{\comskip}[1]{#1}

\newcommand{\rjh}[1]{\textcolor{red}{#1}}
\newcommand{\rjhc}[1]{\comskip{\textcolor{red}{\texttt{#1}}}}

\definecolor{mygreen}{rgb}{0.0,0.4,0.25}
\newcommand{\rlj}[1]{{\color{blue}#1}}

\title{Giant leaps and long excursions: fluctuation mechanisms in systems with long-range memory}
\author{Robert L. Jack}
\affiliation{Department of Applied Mathematics and Theoretical Physics, University of Cambridge, Wilberforce Road, Cambridge CB3 0WA, United Kingdom}
\affiliation{Department of Chemistry, University of Cambridge, Lensfield Road, Cambridge CB2 1EW, United Kingdom}
\author{Rosemary  J. Harris}
\affiliation{School of Mathematical Sciences, Queen Mary University of London, Mile End Road, London E1 4NS, United Kingdom}
\begin{abstract}
We analyse large deviations of time-averaged quantities in stochastic processes with long-range memory, where the dynamics at time $t$ depends itself on the value $q_t$ of the time-averaged quantity.    First we consider the elephant random walk and a Gaussian variant of this model, identifying two mechanisms for unusual fluctuation behaviour, which differ from the Markovian case.  In particular, the memory can lead to large-deviation principles with reduced speeds, and to non-analytic rate functions.  We then explain how the mechanisms operating in these two models are generic for memory-dependent dynamics and show other examples including a non-Markovian symmetric exclusion process.
\end{abstract}

\maketitle

\section{Introduction}

Memory effects and long-range temporal correlations are important in many physical systems~\cite{KeimNagel,Scalliet2019-memory,Kappler2018,Mieghem2013,Zhang2019}, and in other scientific fields ranging from biology to telecommunications to finance~\cite{Rangarajan03,Beran13}.  
It is particularly notable that long-ranged memory can change fluctuation behaviour qualitatively, compared to Markovian (memory-less) cases.
Demonstrations of this include non-Markovian random walks~\cite{Trimper2004,Hod2004,Baur2016,Budini2016,Budini2017,Rebenshtok2007}, models of cluster growth~\cite{Klymko2017,kggw18,Jack2019-growth}, and agent-based models where decisions depend on past experience~\cite{Harris2015-peak-end}.  The distinction between Markovian and non-Markovian systems is also important when formulating general theories.  For example, a  large deviation theory of dynamical fluctuations is now established for Markovian systems~\cite{denH-book,Lecomte2007,Derrida2007,Touchette2009,Chetrite2015,Jack2019ergo}, but memory can lead to new effects which cannot be captured by the standard theory~\cite{Harris2009,Harris2015,Maes2009,Fagg-semi-arxiv,Jack2019-growth}.  In particular, one finds~\cite{Harris2009,Harris2015} a breakdown of the standard large-deviation principles (LDPs) that hold quite generically in finite Markovian systems~\cite{Chetrite2015}.

In this work, we consider non-Markovian systems where the dynamics depend explicitly on a time-averaged current, whose value at time $t$ is denoted by $q_t$.   This is a simple type of memory that occurs in a wide range of physical models~\cite{Trimper2004,Harris2015-peak-end,Budini2017,Jack2019-growth,Harris2009,Harris2015}.  Using methods of large-deviation theory~\cite{denH-book,Lecomte2007,Derrida2007,Touchette2009,Chetrite2015,Franchini2017}, we show how this long-range memory can lead to anomalous fluctuations of $q_t$.  We explain that much of this behaviour can be understood by considering two generic fluctuation mechanisms, where memory plays an intrinsic role.
These general mechanisms are useful for classifying previous results for non-Markovian systems, and for identifying new phenomena.

We illustrate these mechanisms by analysing current fluctuations in  the elephant random walk (ERW) of~\cite{Trimper2004,Baur2016,Bercu_2017,kenkre}, and a related process which we call the Gaussian elephant random walk (GERW).
A key difference from Markovian systems is that large (rare) fluctuations in these models are associated with currents that are strongly time-dependent~\cite{Harris2009,Harris2015,Jack2019-growth,Franchini2017} -- a large current at early times biases the subsequent evolution and can
trigger anomalous fluctuations that persist for large times.  The two specific mechanisms that we discuss are: (i) a very large initial current flow in a finite time interval, which results in anomalously large deviations
(specifically, an LDP for $q_t$ with a speed that is less than $t$~\cite{Harris2009,Harris2015}); and (ii) a large initial current that occurs over a sustained time interval, which leads to a breakdown of the central limit theorem (CLT) for $q_t$~\cite{Trimper2004} and an LDP which generically has a non-analytic rate function~\cite{Jack2019-growth}.  
{The GERW illustrates} mechanism (i), which we refer to as an \emph{initial giant leap} (IGL); the ERW illustrates
mechanism (ii) which we {call} a \emph{long initial excursion} (LIE).  We also describe several other examples of systems in which these mechanisms occur.

The structure of the paper is as follows.  Sec.~\ref{sec:models} introduces the models that we analyse, and some relevant theory.  In Sec.~\ref{sec:results-ele} we describe the large-deviation behaviour of the ERW and GERW models.   Sec.~\ref{sec:mech-general} gives a general theory for the IGL and LIE mechanisms and Sec.~\ref{sec:example-results} describes how this theory plays out in several other models, to illustrate its applicability.  Sec.~\ref{sec:outlook} gives a summary of the main conclusions and open questions.  Additional details of calculations are given in Appendices.


\section{Models and methods}
\label{sec:models}

\subsection{Definitions of ERW and GERW}

The ERW is a random walk model, in discrete time~\cite{Trimper2004,Hod2004}.  The  position of the elephant after step $t$ is $x_t$.  In the variant of the model that we consider here, $x_t$  takes values in a finite domain $\{0,1,\dots,L-1\}$, with periodic boundaries.    (The choice of periodic boundaries does not change the physical behaviour but it is useful when comparing the large-deviation behaviour with that of Markovian systems, see Sec.~\ref{sec:memory-discuss}.  In the following we do not distinguish our periodic variant from the original ERW, except in the rare cases where this is necessary.)
We take $x_0=0$ and denote the displacement of the elephant {on} step $t$ by $\Delta x_{t}$.
Hence the time-averaged current is 
\beq
q_t=\frac{1}{t} \sum_{\tau=1}^t \Delta x_\tau 
\label{equ:q-ele}
\eeq
with $q_0=0$.   The dynamical rule is that 
\beq
\Delta x_t=\pm1 \quad\hbox{ with probability } \quad \frac{1\pm a q_{t-1}}{2} \;,
\label{equ:ele-rule}
\eeq 
where $a \in (-1,1)$ is a parameter that corresponds to $2p-1$ in~\cite{Trimper2004}.  

The GERW is similar to the ERW, except that the position $x_t$ is a real number in $[0,L)$, still with periodic boundaries.  The dynamical rule is that $\Delta x_t$ is a Gaussian-distributed real number with mean $aq_{t-1}$ and variance unity.   Hence the mean value of $\Delta x_t$ (conditional on $q_{t-1}$) is the same as for the ERW. {At first glance, one might also expect fluctuations in the two models to behave similarly but in fact their memory-induced large-deviation behaviour is very different.  Note that although the GERW is rather simple to analyse, it is useful to study in detail as a contrast to the ERW and to illustrate the strong effects of memory.}

\subsection{Large deviations for Markovian and non-Markovian dynamics}
\label{sec:memory-discuss}

For the ERW and GERW, we consider the probability density for $q_t$ at large times, which we denote by $p_t(q)$. 
We will be {particularly} concerned with the tails of this probability distribution and
the associated fluctuation mechanisms, which are characterised by \emph{large deviation theory}~\cite{denH-book,Lecomte2007,Derrida2007,Touchette2009,Chetrite2015}.  This theory describes rare fluctuations, outside the range of CLTs and their generalisations~\cite{Baur2016}.
In recent years, it has been applied to time-averaged quantities in many physical systems, yielding important new insights~\cite{Lebowitz1999,Derrida2007,Garrahan2007,Hedges2009,Gingrich2016}. 

For finite Markov chains, there is a well-established large deviation theory due originally to Donsker and Varadhan (DV)~\cite{DonVarI,DonVarII,DonVarIII,DonVarIV}, see for example~\cite{Touchette2018,Jack2019ergo} for recent summaries. 
 Within this theory,
time-averaged quantities such as $q_t$ obey LDPs of the form
\beq
p_t(q) \simeq \exp[ -t I(q) ] ,
\label{equ:ldp-std}
\eeq
where $t$ is called the \emph{speed} of the LDP, and $I$ the \emph{rate function}.  
More generally, one may also consider
LDPs of the form
\beq
p_t(q) \simeq \exp\left[ -t^\theta I(q) \right] .
\label{equ:ldp-theta}
\eeq
{with $\theta \neq 1$.  In some of the non-Markovian models considered here we find $0<\theta<1$ so the speed $t^\theta$ is reduced, compared to the Markovian case.}   Physically, this means that the memory makes large fluctuations less rare~\cite{Harris2009,Harris2015}.
See~\cite{Nickelsen2018,Gradenigo2019,Meerson2019} for some other examples where LDPs with reduced speed are associated with enhanced fluctuations.

For large deviations, an important quantity is the cumulant generating function for $q_t$:
\beq
G(\lambda,t) = \log \langle \ee^{\lambda t q_t} \rangle \; .
\label{equ:cgf-G}
\eeq
To analyse the limit of large $t$, we consider the scaled cumulant generating function (SCGF) which can be defined generally for LDPs with speed $t^\theta$:
\beq
\psi_\theta(\lambda) = \lim_{t\to\infty} \frac{1}{{t^\theta}} \log \langle \ee^{\lambda t^{\theta} q_t} \rangle \; . 
\label{equ:def-psi}
\eeq
For the usual case $\theta=1$ we omit the subscript and write simply $\psi(\lambda)$.
If the limit (\ref{equ:def-psi}) exists and certain other technical conditions are met then the Gartner-Ellis theorem states that the LDP (\ref{equ:ldp-theta}) holds with
\beq
I(q) = \sup_\lambda [ \lambda q - \psi_\theta(\lambda) ] \; .
\label{equ:I-sup}
\eeq

The classical (DV) theory deals with LDPs of speed $t$.  Under suitable assumptions, the rate function can be shown to be analytic and strictly convex.  
For processes on discrete state spaces, it is sufficient that (i)~the model is Markovian; (ii)~transition rates (or transition probabilities) are independent of time; (iii)~the model is finite and irreducible (and, for discrete-time systems, aperiodic); (iv) the contribution of each transition to the sum in (\ref{equ:q-ele}) is fully determined by its initial and final state.  For the ERW, condition (i) is violated, but the other conditions still hold.  [The ERW was defined on a finite (periodic) domain so that assumption (iii) is valid.]  This enables a clear comparison with the classical theory.
The striking result of this comparison is that the memory effect in the ERW leads to 
a rate function that is (generically) singular at $q=0$, as we show below. Such behaviour is strictly forbidden in the classical case and is directly attributable to the memory effect, via the breaking of assumption (i).

By contrast, the GERW does not allow such a clear comparison with the classical case.  The model is defined on a compact domain, and assumption (iii) can be generalised to account for this, while still ensuring an analytic rate function.  However, the GERW allows for jumps with $|\Delta x_t|>L$, in which case the contribution to (\ref{equ:q-ele}) is not fully-determined by the initial and final states (due to the periodic boundaries).   In principle, assumption (iv) might be generalised to account for this effect, but the memory effect means that the typical jump size can diverge as $q_t\to\infty$, which would not be allowed in the classical case.   In this sense, the GERW violates the classical assumptions more strongly than the ERW.
We show below that this strong violation can lead to an LDP with reduced speed, $\theta<1$.

In fact, the large-deviation behaviour that we will observe also has implications for a particular class of \emph{Markovian} models.  To see this, note that both the ERW and GERW can be formulated as {Markovian} models for either the current $q_t$ or (equivalently) the displacement $Q_t=tq_t$.  The case of the displacement is more natural: in this case the dynamical rule of the ERW is that $Q_{t+1}=Q_{t}\pm 1$ with probabilities $(1\pm aQ_t/t)/2$.  In this formulation, one sees that the transition probabilities depend explicitly on time.  The  GERW may be formulated in a similar way.   {Representing the models in this way, the Markovian assumption (i) above is now obeyed, but assumption (ii) is violated.}
 It follows that the behaviour presented here can be viewed  as either an extension of the classical theory to a particular class of non-Markovian models, or an extension to a class of Markovian models with explicit time-dependence in the rates.

In terms of methods, it is notable that the SCGF $\psi(\lambda)$ in the classical theory can be characterised as the largest eigenvalue of a matrix, the tilted generator~\cite{Chetrite2015}. 
Hence the rate function is available, via (\ref{equ:I-sup}).  Such a determination of $\psi{(\lambda)}$ is not possible for the models considered here,  and other methods must be used, for example the theory of Dupuis-Ellis~\cite{Dupuis-book} as in~\cite{Franchini2017}, or arguments based on separation of time scales~\cite{Harris2009,Harris2015}.

We close this section by noting that since the (G)ERW models are defined on finite periodic domains, they \emph{cannot} be formulated as Markovian processes for $x_t$, even with time-dependent rates.   The transition rates (or transition probabilities) would need to depend on the winding number around the periodic boundaries.

\subsection{General models}

{Although we use the ERW and GERW as motivating examples, we stress that the mechanisms they illuminate have much wider applicability.  To this end, we introduce here a general notation for describing a broad class of models with similar memory-induced phemenonlogy.}   Several examples are given in Sec.~\ref{sec:example-results}.  

We consider models in which the time $t$ may be continuous or discrete.  Let $\CC_t$ denote the configuration of the (general) model at time $t$, this corresponds to $x_t$ in the (G)ERW.  This $\CC_t$ may come from a finite set as in the periodic  ERW, or it may indicate a vector in some compact domain such as $[0,L]^d$.  The periodic GERW is in this latter class with $d=1$.   

All models considered are jump processes.  
We define a time-averaged quantity that generalises (\ref{equ:q-ele}):
\beq
q_t = \frac{1}{t} \sum_{{\rm jumps}\; j} \alpha_j
\label{equ:q-general}
\eeq
where the sum is over {all jumps up to time $t$} and $\alpha_j$ depends on the properties of jump $j$.  (In the ERW there is one jump on each time step and $\alpha_j=\pm1$ coincides with $\Delta x_t$.)  In discrete time the model is specified by the conditional distribution of $\CC_{t+1}$ given $(\CC_t,q_t)$, supplemented by a rule specifying the contribution $\alpha_j$ for each jump.  
In continuous time the model is specified by a set of jump rates [dependent on $(\CC_t,q_t)$], and a rule specifying the $\alpha_j$.  We assume that the dynamical rules do not depend explicitly on time, but only on $(\CC_t,q_t)$.

All the results that we present can be straightforwardly generalised to the case where $q_t$ is a time-average of a state-dependent quantity (for example $\frac{1}{t}\int_0^{t} b(\CC_{{\tau}}) d{\tau}$ as in~\cite{Jack2019ergo}) but we restrict here to the form (\ref{equ:q-general}).  The models that we consider have scalar $q_t$ but the analysis is easily extended to vectorial $q_t$.
For continuous-time models then some regularisation may be required for (\ref{equ:q-general}) at short times, see for example Sec.~\ref{sec:uni}.  

Consistent with Sec.~\ref{sec:memory-discuss} we observe that these general models can be formulated as Markov processes $(\CC_t,q_t)$ but they are not Markovian for $\CC_t$.  Independent of this mathematical distinction, the physical role of $q$ is to capture the role of memory: its definition depends on the full history of the process.

\section{Fluctuations in the GERW and ERW}
\label{sec:results-ele}

In this section we describe the large-deviation behaviour of (G)ERW models.  For the ERW we draw on results of~\cite{Trimper2004, Franchini2017,Baur2016}  and we characterise the relevant fluctuation mechanisms.
For the GERW then $p_t(q)$ can be computed quite straightforwardly and leads to an LDP with reduced speed, we describe the relevant fluctuation mechanisms in this case too.

\subsection{Preliminary results}
\label{sec:results-prelim}

We summarise here some preliminary results for the ERW and GERW with further detail given in {Appendices~\ref{app:ele-prelim} and~\ref{app:urn}.}
For $0<a<1/2$ the memory is relatively weak in these models and at large times they both obey a CLT, where the variance behaves asymptotically as 
\beq
\langle q_t^2 \rangle  \simeq \frac{1}{t(1-2a)}.
\eeq  
Our work focusses on $a>1/2$ where the memory effect is strong, and both ERW and GERW exhibit superdiffusive behaviour:
\beq
\langle q_t^2 \rangle \simeq \frac{\chi}{ t^{2(1-a)} }
\label{equ:var-a}
\eeq  
where  $\chi$ is an $a$-dependent constant {which we denote by  $\chi_{\rm E}$ and $\chi_{\rm G}$ for the two models.  For the ERW we have $\chi_{\rm{E}}=1/[(2a-1)\Gamma(2a)]$ from~\cite{Trimper2004}.}

For the GERW the total displacement is a sum of Gaussian-distributed increments so $p_t(q)$ is Gaussian at all times (although the increments are neither independent nor identically distributed).
As shown in Appendix A, for large times one has
\beq
p_t(q) \propto \exp\left[- \frac{q^2 t^{2(1-a)} }{ 2\chi_{\rm G} } \right] \;, 
\label{equ:ldp-gauss}
\eeq
{where $\chi_{\rm G}$ can be obtained from the limit of a series solution.}  (The proportionality sign is used because a $t$-dependent normalisation constant has been omitted.  We use this notation in cases where the normalisation is clear from the context.) 
For large times, the corresponding CGF is
\beq
G(\lambda,t) \simeq \frac{\lambda^2t^{2a}\chi_{\rm G}}{2}  \; .
\label{equ:cgf-GERW}
\eeq

We now turn to the ERW.  Since the second derivative of the CGF gives the variance of $tq_t$ (and using that the distribution is symmetric) one obtains from (\ref{equ:var-a}) an expansion in powers of $\lambda$ (at fixed $t\gg1$):
\beq
G(\lambda,t)\simeq \frac{\lambda^2t^{2a}\chi_{\rm E}}{2}  + O(\lambda^4) \; ,
\label{equ:cgf-ERW}
\eeq
 which is similar to (\ref{equ:cgf-GERW}). 
 However, there is no CLT for $q_t$~\cite{Paraan2006,Silva2013} which has consequences for the correction terms in (\ref{equ:cgf-ERW}) and their scaling with $t$.

Baur and Bertoin~\cite{Baur2016} considered typical fluctuations of $q$ at large times (that is, fluctuations with probabilities of order unity).   Their theorem 3 states that $p_t(q)$ is a scaling function of $qt^{1-a}$, as in the GERW.  Hence, $G$ is described by a scaling form at large $t$ 
\beq
G(\lambda,t) \simeq g( \lambda t^a ) \; ,
\label{equ:G-scal-erw}
\eeq
which holds as $t\to\infty$, with the argument of $g$ held fixed.  
In a recent mathematical study, Franchini~\cite{Franchini2017} considered large deviations in P\'olya urn models, which can be mapped onto the ERW~\cite{Baur2016}.
Corollary 12 of~\cite{Franchini2017} establishes that $p_t(q)$ follows an LDP, and that (\ref{equ:G-scal-erw}) extends into the large-deviation regime: taking $t\to\infty$ with fixed $\lambda\ll 1$, one has  
\beq
g(\lambda t^a) \simeq c_{\rm E} t |\lambda|^{1/a}
\label{equ:g-erw}
\eeq 
for some constant $c_{\rm E}$ (dependent on $a$).  This result is discussed in Appendix~\ref{app:urn} and a formula for  $c_{\rm E}$ is given in (\ref{equ:cE}).
Hence for $q\ll 1$ and $t\to\infty$ one has the large-deviation result 
\beq
p_t(q) \propto \exp\left( - \kappa_{\rm E} t |q|^{1/(1-a)} \right)  \; .
\label{equ:ldp-erw-smallq}
\eeq
with $\kappa_{\rm E}= (1-a)(a/c_{\rm E})^{a/(1-a)}$.  
For larger $q$ there are deviations from the scaling form; the full SCGF is given in~(\ref{equ:psi-BB}), as derived in~\cite{Franchini2017}.
Equ.~(\ref{equ:ldp-erw-smallq}) is an LDP with speed $t$ and rate function $I(q)\simeq  \kappa_{\rm E} |q|^{1/(1-a)} $.   As advertised above, the long-ranged memory has resulted in  a rate function that is non-analytic (at $q=0$), except in exceptional cases where $1/(1-a)$ happens to be an even integer.
From a physical perspective, note that if $q_t$ obeyed a CLT then its asymptotic variance would be $1/I''(0)$: here we have $I''(0)=0$, which shows that the scaling is superdiffusive, consistent with (\ref{equ:var-a}).

We emphasise that the distributions (\ref{equ:ldp-gauss},\ref{equ:ldp-erw-smallq}) are sharply peaked as $t\to\infty$. In this sense, both systems are ergodic~\cite{Budini2016}.

\begin{figure}
\includegraphics[width=80mm]{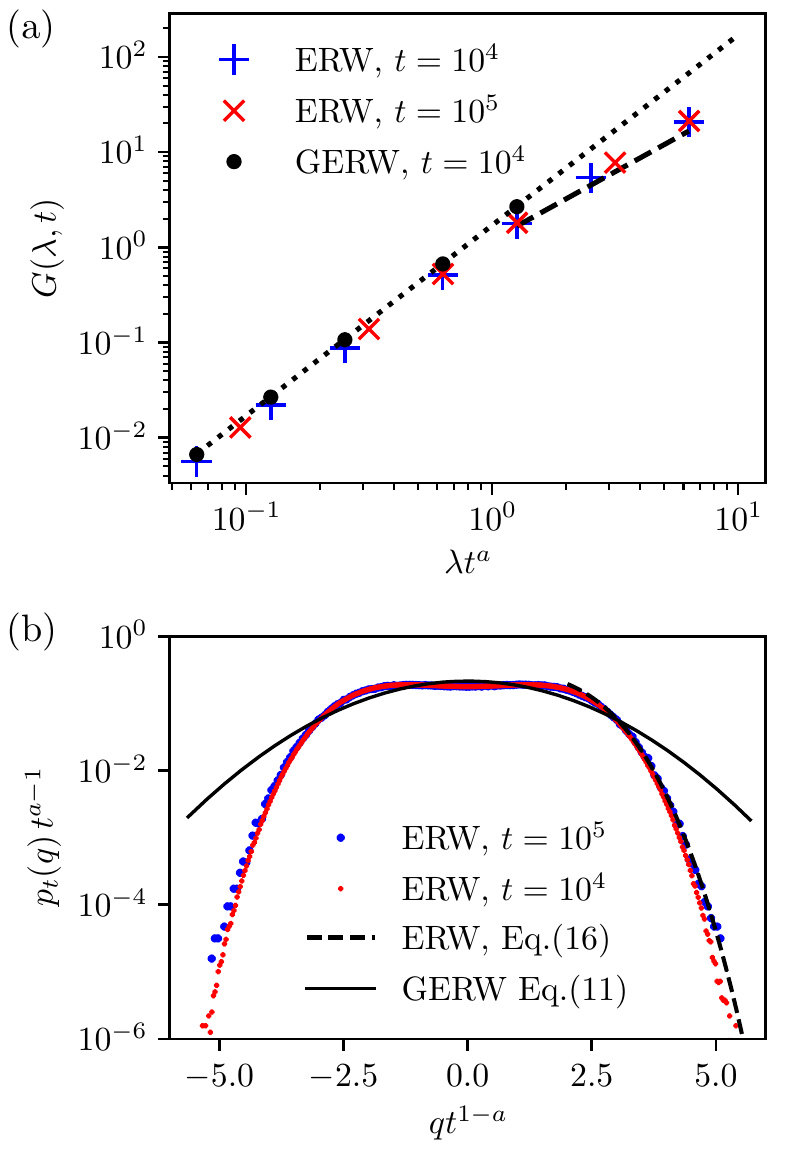}
\caption{{Numerical data for ERW and GERW with $a=0.7$.} (a) CGF for ERW and GERW, plotted as a scaling function of $\lambda t^a$.  The dotted line is the analytical (large-$t$) GERW result (\ref{equ:cgf-GERW}) with $\chi_{\rm G}$ obtained from Appendix~\ref{app:ele-prelim} .  The dashed line corresponds to (\ref{equ:g-erw}) and agrees well with the data, given that it is an asymptotic prediction that assumes that both $t$ and $\lambda t^a$ are large.
There are no fitting parameters.
(b) Distribution of $q$ for the ERW, which collapses to a scaling function of $qt^{1-a}$, as predicted by~\cite{Baur2016}.  (The collapse is not quite perfect, which we attribute to finite-$t$ corrections to scaling.)  The solid line is the analytical (large-$t$) Gaussian distribution of the GERW, for comparison; the dashed line is the prediction (\ref{equ:ldp-erw-smallq}) for the tail of the distribution;
 the constant $\kappa_{\rm E}$ is derived using results from Appendix~\ref{app:urn} while the proportionality constant is determined by fitting to the data.
}
\label{fig:cgf-si}
\end{figure}

Fig.~\ref{fig:cgf-si} shows numerical data, which illustrates these preliminary results. 
For small values of $\lambda t^a$ the CGF for the ERW is proportional to $|\lambda t^a|^2$ and matches the large-$t$ GERW result.  For larger $\lambda t^a$, the CGF for the ERW matches the large-deviation form (\ref{equ:g-erw}) without any fitting [the value of $c_{\rm E}$ is given in (\ref{equ:cE})].   For both ERW and GERW, the distribution $p_t(q)$ is a scaling function of $qt^{1-a}$. The ERW result (\ref{equ:ldp-erw-smallq}) is shown in Fig.~\ref{fig:cgf-si}{(b)} with a dashed line: the value for $\kappa_{\rm E}$ is derived from $c_{\rm E}$ but the proportionality constant in (\ref{equ:ldp-erw-smallq}) is used as a fitting parameter.

 We close this section with a result for conditional averages. 
 The models have fixed initial conditions and averages over the dynamics are denoted by $\langle\cdot\rangle$.  Define also  $\langle\cdot\rangle_{q_\tau}$ as an average that is conditioned on the value of $q_\tau$.  
 For both GERW and ERW, averaging over the possibilities in a single step gives
\beq
(\tau+1) \langle q_{\tau+1}\rangle_{q_\tau} = (\tau + a)  q_\tau \; .
\eeq
It follows~\cite{Trimper2004} that for $t>\tau$, 
\beq
\langle q_t \rangle_{q_\tau} = q_\tau \frac{\Gamma(t+a)\Gamma(\tau+1)}{\Gamma(t+1)\Gamma(\tau+a)} \; .
\eeq
Hence for large $t$ : 
\beq
\langle q_t \rangle_{q_\tau} =  \eta  \frac{ q_\tau  }{t^{1-a}}  \; .
\label{equ:ele-mean}
\eeq
with $\eta = \Gamma(\tau+1)/\Gamma(\tau+a)$.
That is, if the elephant is conditioned to have a non-typical value of $q_\tau$, its subsequent evolution involves regression to the mean (zero) as a power law with exponent $1-a$.  This result will be used in the following to rationalise the large-deviation behaviour of these models.  As a point of comparison, time-averaged quantities in finite Markovian systems generically have power-law relaxation with exponent $1$.

\subsection{IGL mechanism for large deviations in the GERW}
\label{sec:gerw-igl}

We now turn to large deviations, beginning with the GERW.
Consider a discrete-time trajectory with $t$ steps which we represent using its sequence of increments: $\bm{X}=(\Delta x_1,\Delta x_2,\dots,\Delta x_t )$. 
This trajectory occurs with probability $P(\bm{X})$ which is a multivariate Gaussian distribution, so all correlations can be computed exactly (at fixed $t$).
Specifically,
\beq
 P(\bm{X})\propto\exp[-{\cal S}(\bm{X})/2]
\eeq 
 with
\beq
{\cal S}(\bm{X})=\sum_{\tau=0}^{t-1} [ (\tau+1)q_{\tau+1} - q_\tau (\tau+a) ]^2
\label{equ:S-gauss-v1}
\eeq
where we recall that $q_\tau$ is related to the increments $\Delta x$ by (\ref{equ:q-ele}), with $q_0=0$.  

To characterise large deviations, the most likely path that achieves $q_t=q$
can be derived, by conditioning $P(\bm{X})$ on this rare event.  
Collecting terms in (\ref{equ:S-gauss-v1}), one obtains
\begin{multline}
{\cal S}(\bm{X})= 
t^2 q_t^2 + \sum_{\tau=1}^{t-1} q_\tau^2 \left( 2\tau^2 + a^2 + 2 a\tau \right)
\\ - 2 \sum_{\tau=1}^{t-1} q_\tau q_{\tau+1} (\tau+1) \left( \tau + a \right) \; .
\label{equ:S-gauss-v2}
\end{multline}
Conditioning on $q_t$, we arrive at a Gaussian distribution for the $(t-1)$-dimensional vector $\bm{q}=(q_1,q_2,\dots,q_{t-1})$.  This is
\beq
P_{\rm micro}(\bm{q}|q_t) \propto \exp\left(h q_t q_{t-1} -\frac{\bm{q}^T M \bm{q}}{2} \right) \; ,
\eeq 
where $h=t(t+a-1)$, and $M$ is a matrix whose elements can be read from (\ref{equ:S-gauss-v2}).  
The subscript ``micro'' recalls that conditioning on $q_t=q$ is analogous to considering a microcanonical ensemble in thermodynamics.
Completing the square in the exponent, one obtains
\beq
P_{\rm micro}(\bm{q}|q_t) \propto \exp\left[ - \frac{ (\bm{q}-h q_t \bm{\mu})^T M (\bm{q}-h q_t \bm{\mu})}{2} \right] 
\label{equ:Pmic}
\eeq
where $\bm{\mu}$ is given by the $(t-1)$th column of $M^{-1}$.  
Hence the most likely path with $q_t=q$ is given by 
\beq
\langle q_\tau \rangle_{\rm micro} =  \mu_\tau q h \; .
\label{equ:qk-micro}
\eeq
This path depends on the value of $q_t$ and on the associated time $t$.
For finite $t$, the path can be straightforwardly computed numerically (for all $\tau<t$).

\begin{figure}
\includegraphics[width=84mm]{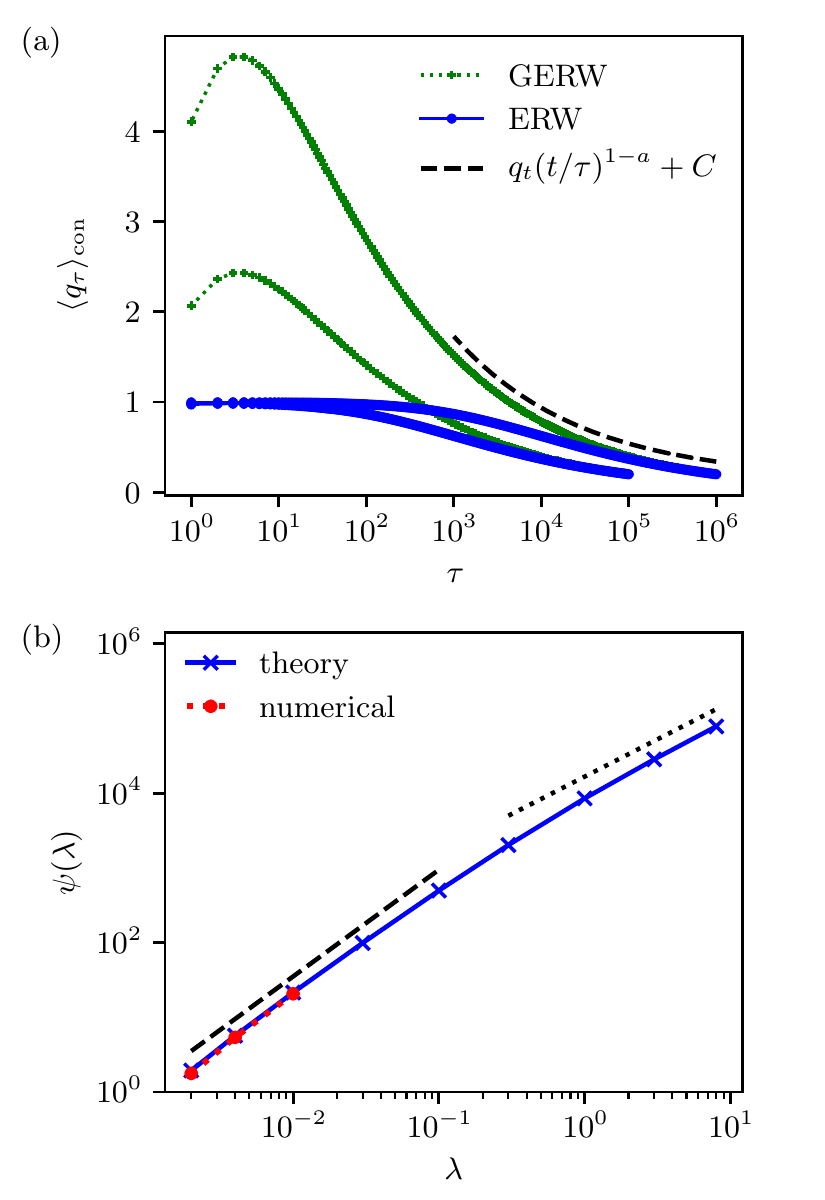}
\caption{%
(a)~Averaged paths $\langle q_\tau\rangle_{\rm con}$ of the controlled dynamics with $t=10^5,10^6$ and
 $\langle q_t\rangle_{\rm con}= 0.2$, for the ERW and GERW with $a=0.7$.  These illustrate the IGL and LIE mechanisms.  
 The dashed line indicates the long-time behaviour (\ref{equ:inst-large-t}) which is common to both ERW and GERW.  (In the plot, this has been offset by $C=0.14$, for clarity.)
(b)~Theoretical estimate for $\psi(\lambda)$ in the ERW {again with $a=0.7$}, derived at $t=10^4$ and compared with numerically exact results for small $\lambda$.
The dashed and dotted lines indicate the predicted power laws $\psi\propto \lambda^{1/a}$ and $\psi\propto \lambda$ respectively.}
\label{fig:inst}
\end{figure}

It is also possible to construct an optimally-controlled process (or auxiliary process) whose typical dynamics generate the most likely path to $q_t=q$.  This is similar to the Doob-transformed dynamics of~\cite{Chetrite2015,Chetrite2015var}, see also~\cite{Maes2008,Simha2008,Simon2009,Jack2010,Jack2015b}
and Appendix~\ref{app:con-gen} of this work for the general theory.  For the GERW, we have derived the optimally-controlled process, see Appendix~\ref{app:control-gerw}.  Its average path is $\langle q_\tau \rangle_{\rm con} = \langle q_\tau \rangle_{\rm micro}$ (for $\tau=1,2,\dots,t$).   
Fig.~\ref{fig:inst}(a) shows results illustrating the average path under the controlled dynamics, also compared with the ERW (see below).  The mechanism for achieving a rare value of $q_t$ is that the GERW makes very large hops on the {first} few steps, after which $q_\tau$ decreases towards $q_t$.  

The results so far are valid for any finite time but we are interested in large deviations as $t\to\infty$.  In this case the problem may be simplified.
 We characterise the most likely path as the minimum of the exponent in (\ref{equ:Pmic}).  Writing the matrix product as a sum over time steps [similar to (\ref{equ:S-gauss-v1})], we fix some $K$ and separate the sum into terms with $\tau\leq K$ and $\tau>K$.  For $\tau>K$ we make the replacement $q_\tau\to \tilde{q}(\tau)$ where $\tilde{q}$ is a smooth function of $\tau$; this allows the sum to be estimated by an integral.   
Fixing values for $q_K$ and $q_t$,  the action can be minimised (exactly) over the function $\tilde q$, which is equivalent to solving the instanton equation in~\cite{Harris2015}.    
One finds 
\beq
\tilde q(\tau) = C_1 \tau^{-a}+ C_2 \tau^{-(1-a)}
\label{equ:inst-gerw}
\eeq 
where $C_1$ and $C_2$ are fixed by the boundary conditions at $\tau=K,t$.  Writing $u=K/t$ (so $0<u<1$), the contribution to ${\cal S}$ from this path is~\cite{Harris2015}
\beq
{\cal S}_1 = ut(2a-1) \frac{(1-u^{2a-1}) \left( q_t-q_Ku^{1-a} \right)^2 }{ (u^{a} - u^{1-a} )^2} \; .
\eeq
For this optimal path (\ref{equ:S-gauss-v1}) then reduces to
\beq
{\cal S} = \sum_{\tau=0}^{K-1} [ (\tau+1)q_{\tau+1} - q_\tau (\tau+a) ]^2 + {\cal S}_1
\eeq
which is to be minimised
over $q_1,q_2,\dots,q_K$.
This is the procedure used to obtain the GERW paths in Fig.~\ref{fig:inst}(a).  We typically take $K=40$, this choice does not strongly affect the results because replacing the discrete sum by an integral is accurate for $K\gg1$.

Remembering that we focus throughout on the case $a>1/2$, the behaviour of (\ref{equ:inst-gerw}) for large $t,\tau$ gives 
\
\beq
\langle q_\tau\rangle_{\rm con}\approx q_t(t/\tau)^{1-a} \; ,
\label{equ:inst-large-t}
\eeq
similar to~\cite{Harris2015}. 
Comparing with (\ref{equ:ele-mean}), we see that the long-time behaviour of the optimally-controlled dynamics matches the natural regression to the mean.

Extrapolating (\ref{equ:inst-large-t}) back to $\tau=1$ indicates that for (rare) paths that end at $q_t$, the first hop should have size $q_t t^{1-a}$, which diverges as $t\to\infty$.  In fact the early-time behaviour is more complex but the size of the first hop is indeed of this order.  The diverging hop is the reason that we call this mechanism an \emph{initial giant leap} (IGL).
It applies in the Gaussian elephant for all fluctuations with $q_t=O(1)$ as $t\to\infty$.

Two comments are in order.  First, the analysis here for $\tau>K$ recovers exactly that of~\cite{Harris2015}, the fact that the distribution of $q_t$ is sharply-peaked under the controlled dynamics can be used to justify the {so-called} temporal additivity assumption in that work, for $\tau\geq K\gg 1$.  However for the early part of the trajectory with $\tau<K$, it is important that the model evolves by discrete time steps and that $q_t$ can change significantly in a single step.  This means that the temporal additivity assumption is not valid in this regime.  For this reason, quantitative results for $p_t(q)$ require a more detailed analysis of early times, without the temporal additivity assumption.  We accomplish this here by analysing numerically the sum of terms with $\tau<K$.  The second comment is that we use the language of a giant leap, but we note that the GERW makes (on average) very large jumps on several of the early time steps.  We explain below that we are using IGL to refer to any divergent displacement $q^*$ in a finite time interval $\tau^*$, see Sec.~\ref{sec:igl-gen}.  

\subsection{LIE mechanism for large deviations in the ERW}
\label{sec:erw-lie}

As discussed in Sec.~\ref{sec:results-prelim}, large-deviation properties of the ERW are available from~\cite{Franchini2017}.  
In particular, there is an LDP for $q_t$ with speed $t$ whose rate function behaves for small $q$ as
\beq
I(q) \simeq \kappa_{\rm E} |q|^{1/(1-a)} \; .
\label{equ:I-erw}
\eeq
Correspondingly,
\beq
\psi(\lambda) \simeq c_{\rm E} |\lambda|^{1/a} \; ,
\label{equ:psi-erw}
\eeq
{for small $\lambda$.} [Recall Equs.~(\ref{equ:g-erw},\ref{equ:ldp-erw-smallq}).]

We characterise here the mechanism responsible for (\ref{equ:psi-erw}), by deriving a controlled process which captures the behaviour of the relevant conditioned path ensemble, see Appendices~\ref{app:con-gen} and \ref{app:control-ele}.  This controlled process is similar to the original process, but now $\Delta x_\tau=\pm1$ with time-dependent probabilities $(1\pm b_\tau)/2$ where $(b_1,b_2,\dots,b_t)$ are variational parameters {that we optimise, to reproduce the large-deviation mechanism.}  

This analysis yields a controlled process for which $\langle q_\tau\rangle_{\rm con}$ is shown in Fig.~\ref{fig:inst}(a): for early times,  typical paths 
have $q_\tau\approx 1$ which is the maximum possible value in the ERW.  This behaviour persists over a finite fraction of the trajectory, which motivates the name, \emph{long {initial} excursion} {(LIE)}.    
For larger times,  $q_\tau$ decreases. 
Fig.~\ref{fig:inst}(b)  shows our theoretical estimate of 
$\psi(\lambda)$ obtained by a variational analysis at finite $t$, compared with numerically exact results from direct simulation.  The theoretical estimate (i) matches the exact result in the region where numerical results are available; (ii) is consistent with (\ref{equ:psi-erw}) for $t^{-a}\ll\lambda\ll 1$; 
(iii) recovers $\psi(\lambda)\simeq |\lambda|$ for large $\lambda$, which is the exact result (since $q\leq 1$).  The controlled dynamics give a good description of the true $\psi(\lambda)$.

{It can also be shown that the averaged paths in Fig.~\ref{fig:inst}(a) capture the true fluctuation mechanism.  We sketch the argument.
At the level of large deviations, the true mechanism is the path measure $P^{\rm con}$ that achieves equality in (\ref{equ:psi-sup}).
From~\cite{Franchini2017}, the large-deviation event $q_t=q$ is associated with a single path, in the sense that the conditional distribution of $q_{\alpha t}$ is sharply-peaked as $t\to\infty$ for all $\alpha\in(0,1]$.  Our ansatz for the controlled process is sufficiently general to capture this path, so minimising over all controlled paths is sufficient to make (\ref{equ:psi-sup}) an equality, and hence to obtain the true mechanism.
This argument also justifies the temporal additivity principle of~\cite{Harris2009} in this case.  

Ref.~\cite{Harris2015} used that principle together with a quadratic expansion of the action about $q=0$, for {(symmetric)} 
models similar to the ERW.   This predicts dominant paths similar to (\ref{equ:inst-gerw}).  The results presented here show that such an expansion is not generically valid: for all $q_t\neq 0$, large-deviation events involve {initial} excursions far from $q_t=0$, and the quadratic expansion breaks down.
Nevertheless, if $q_t\ll1$ then the quadratic expansion is applicable at large times and can be used to show that the optimally-controlled process behaves the same as the GERW for large $\tau$, that is 
$\langle q_\tau\rangle_{\rm con} \approx (t/\tau)^{1-a}q_t\; $ as in (\ref{equ:inst-large-t}). 
 [For the ERW, this result is valid for $a>1/2$ and $q_t\ll1$ with $t,\tau\to\infty$ such that also $\langle q_\tau\rangle_{\rm con} \ll 1$].  A very similar case is analysed in Sec III.C of~\cite{Jack2019-growth}, for a cluster growth model.

Comparing (\ref{equ:inst-large-t}) with (\ref{equ:ele-mean}), we see that the long-time behaviour of the optimally-controlled dynamics matches the natural regression to the mean, for both ERW and GERW.  In other words,  the controlled dynamics is almost that of the original model, when $\tau$ is sufficiently large.
In Sec.~\ref{sec:lie-gen} below, we exploit this fact to show that the scaling $\psi(\lambda)\sim|\lambda|^{1/a}$ of (\ref{equ:psi-erw}) is generic if
optimally-controlled processes have (i)~$\langle q_t\rangle_{\rm con}\approx 1$ until some time $\tau^*\sim t$, and (ii)~$\langle q_\tau\rangle_{\rm con} \sim \tau^{-(1-a)}$ for long times.  This is the sense in which the ERW is a prototype for a general fluctuation mechanism.}

\section{Generic fluctuation mechanisms}
\label{sec:mech-general}

\subsection{Overview of method}

We have explained that the large-deviation behaviour of the ERW and GERW is different from that expected in Markov chains.  Fluctuations in these models occur by mechanisms
where the particle makes a large \emph{excursion} from the origin at early times, which biases all future motion {in the same direction}, via the memory effect.  This leads to a reduced speed in the LDP of the GERW and to a singular rate function in the ERW. The difference between ERW and GERW arises from the different characters of their initial excursions (a giant leap over a finite time for the GERW and a long excursion scaling with trajectory length for the ERW).
In this section we explain that such phenomena are relevant for a broad class of non-Markovian models. 
We provide general conditions under which excursions can occur, and explain their consequences for LDPs.

We consider models where $q_t$ converges to its mean as $t\to\infty$ (to be precise, this is convergence in probability).  We denote this mean value by
\beq
q_\infty = \lim_{t\to\infty} \langle q_t\rangle \; .
\eeq
For simplicity we discuss deviations with $q_t>q_\infty$; the opposite case is a straightforward analogue.
We consider excursions which extend over a time period between $t=0$ and some time $\tau^*$.  
The probability $p_t(q)$ can be bounded from below by restricting to paths where the size of the excursion is at least $q^*$, that is $q_{\tau^*}\geq q^*$.
By conditional probability:
\beq
\log p_t(q) \geq \log p_t(q|q_{\tau^*}\geq q^*) + \log P(q_{\tau^*}\geq q^*)  
\label{equ:cond-bound}
\eeq
where $P(q_{\tau^*}\geq q^*)$ is the probability of the excursion and $p_t(q|q_{\tau^*}\geq q^*)$ is the corresponding conditional probability density for $q_t$.

The inequality (\ref{equ:cond-bound}) is valid for all $q^*,\tau^*$.  Now suppose that  $q,t$ are given and we seek a useful bound on $p_t(q)$: this requires that we choose suitable values for $q^*,\tau^*$.  To this end, we introduce the notation $\langle\cdot\rangle_{q^*,\tau^*}$ for averages that are conditioned on $q_{\tau^*}\geq q^*$. 
Then we choose $q^*,\tau^*$ such that $\langle q_t\rangle_{q^*,\tau^*}=q$ and we further assume that the conditional distribution of $q_t$ is sharply-peaked at this value.  This means that if we consider trajectories where a suitable excursion has already taken place before $\tau^*$, then following the natural dynamics of the model for $t>\tau^*$ will result in $q_t\approx q$ with a probability close to unity.

Under these assumptions, (\ref{equ:cond-bound}) reduces to
\beq
\log p_t(q) \gtrsim \log P(q_{\tau^*}\geq q^*) \; .
\label{equ:pq-bound}
\eeq 
In other words, we now have a more explicit lower bound on $p_t(q)$ which is valid if
\beq
\langle q_t\rangle_{q^*,\tau^*}=q.
\label{equ:q*-q}
\eeq
 (The additional requirement that the conditional distribution is sharply peaked is always obeyed in the following.) 

The strategy in Secs.~\ref{sec:igl-gen} and \ref{sec:lie-gen} below is to characterise situations in which (\ref{equ:pq-bound}) can be used to establish LDPs that differ from those expected in finite Markovian models.  
{In particular, we now} establish a sufficient condition for memory to have a strong effect on the large-$t$ behaviour.  Physically, the idea is that after the excursion, 
  the time-averaged current relaxes to its steady-state value as a power law with exponent $1-a$, as established in (\ref{equ:ele-mean}) for the (G)ERW.
 Finite Markovian systems relax generically as $t^{-1}$, so $a$ encodes the effects of memory, this is related to the fixed-point stability analysis of~\cite{Harris2015}.  The condition that we will require is that for $t>\tau^*$,
\beq
\langle q_t - q_{\infty}  \rangle_{q^*,\tau^*}  \simeq (q^*-q_\infty) {\cal F}(q^*,\tau^*)   \left(\frac{\tau^*}{t}\right)^{1-a}  \; 
\label{equ:vq-a}
\eeq
for some function $\cal{F}$, and some number $a\in(0,1)$.  

We then arrive at the following method for deriving bounds on $p_t(q)$.  We must first establish (\ref{equ:vq-a}) for a particular model, at least for $q^*$ in some range.  To bound $p_t(q)$ 
for specific values of $q,t$, we must then find a combination $q^*,\tau^*$ such that (\ref{equ:vq-a}) holds, with $q_t=q$.  As long as this is possible, the constraint (\ref{equ:q*-q}) is satisfied and the resulting $q^*,\tau^*$ can be substituted into (\ref{equ:pq-bound}) to obtain a bound on $p_t(q)$.  Note that the combination $q^*,\tau^*$ depends in general on $t$; the final step is to take $t\to\infty$ in order to characterise large deviations that occur in this limit. 
This strategy is similar to those used in~\cite{Jack2019-growth}.

\subsection{Generic IGL mechanism} 
\label{sec:igl-gen}

We now show how a generic IGL mechanism leads to a useful bound.  We achieve this by laying out the properties that a model should have, 
in order that this mechanism is relevant.  A defining feature of the IGL is that it takes place over a finite time period $\tau^*$ and that the size of the excursion diverges in the limit $t\to\infty$.  

The first requirement is that the model of interest supports very large excursions.
To characterise their probability,
we require that there exists some $\tau^*$ such that for $q^*\to\infty$ we have
\beq
\log P(q_{\tau^*}\geq q^*)\simeq -\gamma |q_*-q_\infty|^\beta \; ,
\label{equ:exc-IGL}
\eeq 
with $\gamma,\beta>0$.  
Since we consider divergent excursions, we require
that (\ref{equ:vq-a}) remains valid even as $q^*\to\infty$.  In the following we take $\tau^*$ to be a fixed parameter, the choice of its value is discussed below.  We define
\beq
f_*(\tau^*) = \lim_{q^*\to\infty} {\cal F}(q^*,\tau^*)
\label{equ:f*}
\eeq
which we require to be strictly positive.  These requirements place strong constraints on the range of models for which the IGL mechanism will determine the large deviations but, as we demonstrate, such models do indeed exist.
Then (\ref{equ:cond-bound},\ref{equ:vq-a}) with $q=\langle q_t\rangle_{q^*,\tau^*}$ yield
\beq
-\log p_t(q) \lesssim  t^{\beta(1-a)}   |q-q_\infty|^\beta \kappa_{\rm IGL} 
\label{equ:rho-IGL}
\eeq
with $\kappa_{\rm IGL}=\gamma f_*(\tau^*)^{-\beta} \tau_*^{-\beta(1-a)}$.  

Equ.~(\ref{equ:rho-IGL}) corresponds to an LDP with speed $t^{\beta(1-a)}$.
If this speed  is less than $t$, fluctuations are qualitatively larger than one finds in generic Markovian systems.
In principle the bound (\ref{equ:rho-IGL}) can be optimised over $\tau^*$.
However, (\ref{equ:rho-IGL}) already establishes that the speed of the LDP {can be} less than $t$, without any requirement for optimisation over $\tau^*$.  This is the central result.
In this sense, the specific value of $\tau^*$ is not crucial.

 The GERW satisfies all the requirements for the IGL mechanism, with $\beta=2$; one may take $\tau^*=1$.  
The applicability of (\ref{equ:vq-a}) was already shown in (\ref{equ:ele-mean}).  {The resulting bound} is consistent with the exact result (\ref{equ:ldp-gauss}), it gives the right scaling with $t$ and the correct general mechanism.  However, the constant {$\kappa_{\rm IGL}$} obtained from this generic argument does not coincide with the prefactor in {the exponent of} (\ref{equ:ldp-gauss}): obtaining that result requires the more detailed (model-dependent) calculation of Sec.~\ref{sec:gerw-igl}. 

 We note in passing that some arguments of~\cite{Harris2015} are similar to those of this section, but the connection between the giant leap and the reduced speed of the LDP was neglected in that work.  In particular, the requirement that (\ref{equ:vq-a}) must hold as $q^*\to\infty$ means that some care is required when applying the arguments of~\cite{Harris2015} to generic models; {they are not valid in the ERW, for example.}

\subsection{Generic LIE mechanism} 
\label{sec:lie-gen}

The LIE mechanism is generically associated with excursions that have finite $q_*$ but diverging $\tau_*$ (proportional to $t$).
This may be compared with the IGL, which has fixed $\tau^*$ and diverging $q^*$.
The LIE mechanism has two central requirements, which must hold for some $q^*$, different from $q_\infty$.  First,~(\ref{equ:vq-a}) must hold asymptotically for $1\ll \tau^*\ll t$. Second,
\beq
f_\ddag(q^*) = \lim_{\tau^*\to\infty}{\cal F}(q^*,\tau^*)
\label{equ:f-ddag}
\eeq
must be strictly positive.  Comparing with (\ref{equ:f*}), the roles of $q^*,\tau^*$ are reversed.  

Under these conditions, we assume that there is an LDP with speed $t$ as in (\ref{equ:ldp-std}), verify the self-consistency of this assumption, and establish a bound on the rate function {$I(q)$} for $|q-q_\infty| \ll 1$.  
Since $\tau^*$ is proportional to $t$, this means that 
\beq
P(q_{\tau^*}\geq q^*) \simeq \exp[-\tau_* I(q_*)]
\label{equ:exc-LIE}
\eeq
which is analogous to (\ref{equ:exc-IGL}).  Using this with (\ref{equ:pq-bound},\ref{equ:vq-a}) yields
\beq
-\log p_t(q) \lesssim  t \kappa_{\rm LIE} |q-q_\infty|^{1/(1-a)} 
\label{equ:rho-LIE}
\eeq
with
\beq
\kappa_{\rm LIE} =  I(q^*)  \left(\frac{1}{|q_*-q_\infty| f_\ddag(q^*)}\right)^{1/(1-a)} \; .
\eeq
The result (\ref{equ:rho-LIE}) is consistent with the assumption of an LDP with speed $t$, but it shows (for $a>1/2$) that the rate function increases from zero more slowly than any quadratic function.
As noted above, this means that $I''(0)=0$,  corresponding to superdiffusive scaling.

In addition, by Varadhan's lemma [a standard result in large deviation theory~\cite{Touchette2009,denH-book}, which amounts to the inverse Legendre transform of (\ref{equ:I-sup})], one obtains 
\beq
\psi(\lambda) {\gtrsim} \sup_q [\lambda q - |q-q_\infty|^{1/(1-a)} \kappa_{\rm LIE} ]
\eeq 
which gives
\beq
\psi(\lambda) \gtrsim \lambda q_\infty + |\lambda|^{1/a} c_{\rm LIE}
\eeq
with 
\begin{align}
c_{\rm LIE} & = a\left( \frac{1-a}{\kappa_{\rm LIE}} \right)^{(1/a)-1} \; .
\end{align}
All these generic arguments are consistent with the behaviour of the ERW, which has $q_\infty=0$.  In particular, the requirement {for (\ref{equ:vq-a}) to hold asymptotically} follows from (\ref{equ:ele-mean}).

\begin{figure}
\includegraphics[width=70mm]{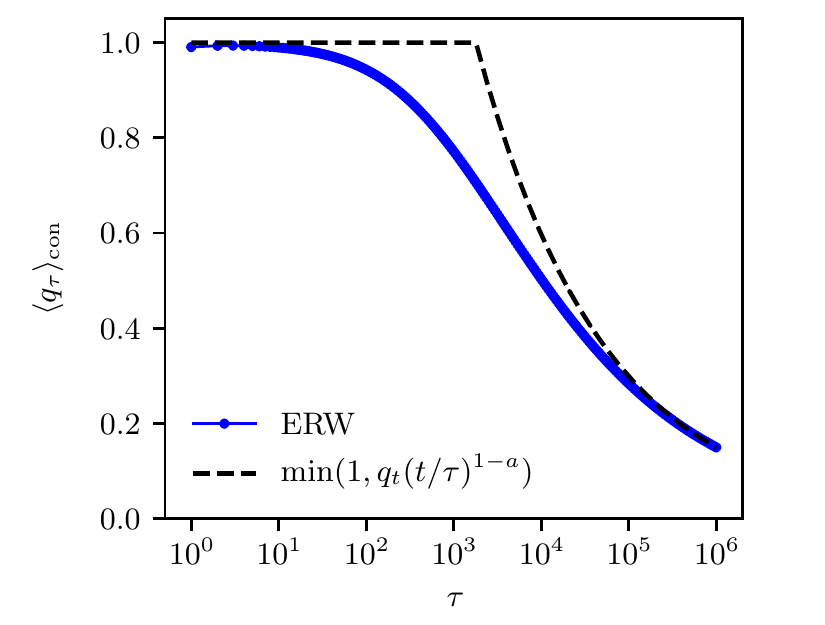}
\caption{Comparison between the optimally controlled path for the ERW (similar to Fig.~\ref{fig:inst}) and the corresponding  generic LIE path used to derive (\ref{equ:rho-LIE}).  We take $a=0.7$ with 
$q_t=0.15$ and $t=10^6$.  The generic LIE path has an excursion with $q^*=1$, after which $q_\tau$ relaxes back towards zero, as the system follows its natural dynamics (\ref{equ:ele-mean}). The generic LIE path does not capture the details of the optimally-controlled (instanton) path which means that the coefficient $\kappa_{\rm LIE}$ does not match $\kappa_{\rm E}$ in (\ref{equ:ldp-erw-smallq}), but the generic LIE argument is sufficient to capture the non-quadratic form of the rate function at $q=0$.}
\label{fig:LIE-path}
\end{figure}

Moreover, the results of Sec.~\ref{sec:erw-lie} indicate that the true fluctuation mechanism for the ERW is an LIE with $q^*=1$.  Also $p_t(1) = (1+a)^{t-1}/2^t$ because all hops have $\Delta x_t=1$ in this case, so $I(1) = \log[2/(1+a)]$. 
 The coefficient in (\ref{equ:ele-mean}) is $\eta\simeq\tau^{1-a}$ as $\tau\to\infty$, which means $f_\ddag(q^*)=1$.  
 Hence the bound (\ref{equ:rho-IGL}) holds with $\kappa_{\rm LIE} = \log[2/(1+a)]$.  The exact result for the ERW can be obtained from $\kappa_{\rm E}= (1-a)(a/c_{\rm E})^{a/(1-a)}$ as quoted in Sec.~\ref{sec:results-prelim}, together with (\ref{equ:cE}).  
 
 For the representative case $a=0.7$, we find $\kappa_{\rm E}=0.04$ while $\kappa_{\rm LIE}=0.16$.  Given that the generic LIE argument is much simpler than the full calculation of $\kappa_{\rm E}$, this level of agreement is reasonable.  
 The generic LIE argument is based on a simple path (or equivalently a simple controlled process) that includes a long excursion: the path 
 is illustrated in Fig.~\ref{fig:LIE-path}, where it is compared with the optimal LIE path discussed in Sec.~\ref{sec:erw-lie}.  The generic LIE path captures the correct qualitative behaviour and matches the optimal path for small and large times.  However, the agreement is not quantitative, and the difference between $\kappa_{\rm LIE}$ and $\kappa_{\rm E}$ reflects this.

\subsection{Discussion of generic IGL and LIE mechanisms}

We summarise the difference between the IGL and LIE mechanisms. The IGL makes a giant (divergent) excursion in a finite time and leads to an LDP with reduced speed.  The LIE makes a finite excursion over a long (divergent) time period; it leads to an LDP with speed $t$, and 
to a rate function with $I''(q_\infty)=0$ which is (generically) non-analytic at $q_\infty$.
In all the examples that we have managed to construct, the IGL mechanism relies on microscopic transition rates that diverge as $q_t\to\infty$, in order to satisfy (\ref{equ:vq-a}).

The IGL mechanism has an interesting analogy with condensation in interacting-particle systems~\cite{Grosskinsky03,Evans14b}: to achieve $q_t=q$ the system must support an
excess current which may be distributed over {a macroscopic fraction} of the time period (as in the LIE), or \emph{condensed} into a finite time interval (the IGL).
A similar phenomenon is described by the ``single-big-jump'' principle for sums of random variables (including certain types of correlated process)~\cite{Vezzani19}; the particular history-dependence in our models, with $a>0$, constrains the condensation to take place at the beginning of the time period.

We close this section by noting that (\ref{equ:rho-IGL},\ref{equ:rho-LIE}) are both lower bounds on the probability $p_t(q)$.  Physically, this means that fluctuations \emph{can} take place by IGL and LIE mechanisms, so fluctuations of a given size $q$ are \emph{at least as likely} as (\ref{equ:rho-IGL},\ref{equ:rho-LIE}) predict.  We have not ruled out competing mechanisms that might allow fluctuations of the same size to occur {in a  more likely way}.  As a simple example, an LIE bound can be obtained for the GERW but does not accurately describe the probability of rare fluctuations, because the IGL mechanism is available and occurs with (much) higher probability.  [Indeed it is easy to see that the IGL mechanism, if available, will always dominate the LIE mechanism if $\beta(1-a) < 1$.] To rule out competing mechanisms, one would need a matching upper bound on the probability; this seems to require more detailed (model-dependent) analysis.

\section{Example models exhibiting IGLs and LIEs}
 \label{sec:example-results}

 By considering IGLs and LIEs, we have established simple and generic requirements which enable bounds on the probabilities of large-deviation events.  It is straightforward to construct (or identify) other models that exhibit these mechanisms.  In this section we give a brief discussion of three such cases.  Similar to the ERW in Sec.~\ref{sec:erw-lie}, we establish bounds on the probabilities of large excursions by using arguments based on optimal control theory, these computations then enable us to check conditions for the IGL and LIE mechanisms.  Our main purpose here is not to describe the model behaviour in detail, but rather to illustrate the general relevance of the identified mechanisms.
 
\subsection{IGL in unidirectional hopping model}
\label{sec:uni}

As an example in continuous time, we modify the unidirectional walker model of~\cite{Harris2009}.  
Similarly to the ERW, we consider a particle with integer-valued position $x_t$ which we identify with the configuration $\CC_t$.  The
particle always hops in the same direction so $\Delta x_t=1$.  We define $q_t$ as the total {time-averaged} displacement which corresponds to (\ref{equ:q-general}) with $\alpha_j=1$ for all jumps.  In the variant of the model that we consider, the particle makes its first jump at time $t_0$; subsequent jumps occur with rate 
\beq
r(q_t)=a q_t \; ,
\label{equ:v-uni}
\eeq  
{where $0 < a < 1$.}
The regularisation parameter $t_0$ is important because if one allows jumps to occur at arbitrarily early times then $q_t$ in (\ref{equ:q-general}) can become arbitrarily large after just one jump; combined with (\ref{equ:v-uni}), this can lead to pathological fluctuations.

The results of~\cite{Harris2009} indicate that large deviations with $q_t>0$ involve a giant leap of size $q^*\sim t^{1-a}$, leading to an LDP with speed $t^{1-a}$.  However, that work made an assumption of temporal additivity which (strictly-speaking) is valid only for $t_0\gg1$.  Here we discuss the case where $t_0$ takes any positive value; we show that the IGL mechanism operates, and $p_t(q)$ can be bounded as in (\ref{equ:rho-IGL}), which is consistent with an LDP with speed $t^{1-a}$.

To analyse the IGL we take $\tau^*=2t_0$. In this case
we show in Appendix~\ref{app:uni} that 
\beq
\log P(q_{\tau^*}\geq q^*) \gtrsim - \gamma_{\rm uni} q^* t_0 \; ,
\label{equ:uni-P-bound}
\eeq
with $\gamma_{\rm uni}=O(1)$ as $q^*\to\infty$.
That is, the probability of a large excursion to $q^*$ in a finite time decays at most exponentially in $q^*$.
This establishes the requirement (\ref{equ:exc-IGL}) for an IGL.

Moreover, after the excursion the average displacement obeys
\beq
\tau \frac{\partial}{\partial \tau} \langle q_\tau \rangle_{q^*, {\tau^*}} = (a-1)\langle q_\tau \rangle_{q^*,\tau^*} \; ,
\label{equ:dotq-uni}
\eeq
which follows directly from the master equation of the model.  Similar to (\ref{equ:ele-mean}),
integrating this equation yields $\langle q_\tau \rangle_{q^*,\tau^*}=q^*(\tau_*/\tau)^{1-a}$ which is exactly the required condition (\ref{equ:vq-a}) with $q_\infty=0$ and ${\cal F}(q^*,\tau^*) = 1$.  Note that this holds even as $q^*\to\infty$, which is related to the fact that ${r}(q^*)$ diverges in this limit.   
To apply (\ref{equ:pq-bound}) requires that 
the conditional distribution of $q_t$ after the excursion is sharply-peaked: this is easily verified.

Hence, the conditions for an IGL are in place and we have established (\ref{equ:rho-IGL}) with $\beta=1$ and $f_*{(\tau^*)}=1$, that is
\beq
-\log p_t(q) \lesssim  t^{1-a} \kappa_{\rm uni} q   \; ,
\label{equ:uni-final}
\eeq
with $ \kappa_{\rm uni} ={2^{a-1}}  \gamma_{\rm uni} t_0^a $, using $\tau^*=2t_0$, from above.
This corresponds to an LDP with speed $t^{1-a}$ as shown in~\cite{Harris2009,Harris2015} by arguments based on an assumption of temporal additivity.
Our analysis avoids any such assumption; it also shows that the unusual speed of the LDP arises because the fluctuation mechanism is an IGL.

 The result (\ref{equ:uni-final}) applies to the unidirectional model with ${r}(q)=aq$ but, in fact, the main ingredient required in the analysis was  $\lim_{q\to\infty} [{r}(q)/q]=a$ (with $0<a<1$).  We therefore expect the IGL mechanism to operate for a broad class of models where this assumption holds. 
 
\subsection{LIE in cluster growth models}
\label{sec:growth}

We consider a model of a growing cluster as in~\cite{Klymko2017,kggw18,Jack2019-growth}.  The cluster contains two types of particles (for example, red and blue) whose numbers at time $t$ are $n^R_t$ and $n^B_t$.  The cluster evolves in discrete time and a single particle is added on each step, so $n^R_t+n^B_t=t$.  (This is the irreversible model of~\cite{Klymko2017}, in that particles are added but never removed.)  The configuration is given by $\CC_{{t}}=(n^R_{{t}},n^B_{{t}})$ and we take $q_t=(n^R_t-n^B_t)/t$ which means that $\alpha_t=\pm1$ in (\ref{equ:q-general}) according to whether  a red or blue particle is added. 

On step $t$, the added particle is red ($+$) or blue ($-$) with probability $(1\pm\tanh Jq_{t-1})/2$ where $J>0$ is a parameter that reflects the difference in energy on adding either a red or blue particle.  In this case, the dynamics of the quantity $m_t=(n^R_t-n^B_t)$ is similar to that of the ERW position $x_t$, but with the nonlinear tanh function replacing the linear function in (\ref{equ:ele-rule}).  This nonlinearity leads to a symmetry-breaking transition: for $J<{1}$ then $q_t\approx 0$ at long times (``mixed'' clusters) but  for $J>1$ then $q_t\approx \pm \bar{m}$, which corresponds to spontaneous de-mixing. Here $ \bar{m}$ is the order parameter for the underlying phase transition~\cite{Klymko2017}. 
 Large deviations in this model were discussed previously in~\cite{kggw18,Jack2019-growth}, it may be also formulated as an urn model so the results of~\cite{Franchini2017} are applicable.

In the mixed (one-phase) regime, the behaviour of this model is qualitatively similar to the ERW.  It can be analysed similarly to Sec.~\ref{sec:erw-lie}, using the same (general) controlled model: red/blue particles are added with probabilities $(1\pm b_t)/2$. The  theoretical arguments of Appendix~\ref{app:con-gen} can then be applied.  Indeed, these ideas were already applied to the growth model in~\cite{Jack2019-growth}: for $1/2<J<1$ this led to a result analogous to (\ref{equ:cgf-ERW}), with $a=J$.  However, that paper did not come to a definitive conclusion about the speed of the LDP in this regime.  The general results of the present work can be used to resolve this open question, and to understand the rare-event mechanism.  We outline the argument below (again for $1/2<J<1$).

The results of~\cite{Franchini2017} prove that the LDP for this model must have speed $t$, so one may expect an LIE mechanism, similar to the ERW.  Moreover, Ref.~\cite{Jack2019-growth} showed that (\ref{equ:ele-mean}) holds in this system for relaxation as $t\to\infty$ after an initial excursion.  However, contrary to the ERW, this result is now valid only for $\langle q_t\rangle_{q_\tau}\ll1$.  This establishes that (\ref{equ:vq-a}) holds, but only for small values of $q^*$.  

We therefore fix some small value for this parameter and construct the LIE, using (\ref{equ:vq-a}) as in Sec.~\ref{sec:lie-gen} to fix $\tau^* =  t(q^*f_\ddag(q^*)/q_t)^{-1/(1-a)} $ so that the natural dynamics after the excursion arrives at $q_t$ with probability 1.  By~\cite{Franchini2017}, this excursion has ${\rm Prob}(q_{\tau^*}\geq q^*) \simeq \exp[-\tau^* I(q^*)]$, although the rate function $I$ is not known explicitly.
These results can be used with (\ref{equ:pq-bound}) to obtain 
\beq
-\log p_t(q) \lesssim  t \kappa_{\rm LIE} |q|^{1/(1-a)}  \; ,
\label{equ:rho-clust}
\eeq
as in (\ref{equ:rho-LIE}).
Since the validity of (\ref{equ:vq-a}) is restricted to small $q^*$, this construction is restricted to small $q$ (strictly positive and fixed as $t\to\infty$).  Still, this generic bound is sufficient to establish the non-analytic behaviour of the rate function at $q=0$.  The use of a fixed small value of $q^*$ is convenient for this argument but is not expected to be optimal for the large-deviation mechanism; in fact we anticipate the true large-deviation mechanism to involve an excursion with $q^*=1$, as for the ERW.  This means that the prefactor $\kappa_{\rm LIE}$ is likely to be far from optimal, but the scaling (\ref{equ:rho-clust}) is expected to be robust.

The overall picture is that for small values of $q_t$ (fixed as $t\to\infty$), the cluster growth model with $1/2<J<1$ behaves similar to an ERW with $a=J$, exhibiting an LIE fluctuation mechanism and a rate function that increases from zero with exponent $1/(1-a)$.  Physically, the similarity can be explained by an argument similar to the fixed-point stability analysis of~\cite{Harris2015}, because the exponent that appears in the LIE bound only depends on the asymptotic (long-time) dynamics close to the fixed point.  For models that can be formulated as urns~\cite{Franchini2017}, we therefore expect these similarities with the ERW to be generic, based on an expansion of the urn function about the fixed point. 

\subsection{LIE in a non-Markovian exclusion process}
\label{sec:esep}

We consider a non-Markovian symmetric exclusion process {(SEP)} where $N$ particles hop {in continuous time} on a periodic one-dimensional lattice of $L$ sites, {subject to the constraint that each site may contain at most one particle.  We define} $n_i=1$ if site $i$ contains a particle, and $n_i=0$ otherwise.  A configuration is specified as $\CC=(n_1,n_2,\dots,n_L)$.  The time-averaged current is $q_t=(Lt)^{-1}\sum_{{\rm jumps}\, j}\Delta x_j$, as in (\ref{equ:q-general}), where the sum is over all particle hops, with $\Delta x_j=\pm1$ according to whether the hop is to the right or the left. 
Large deviations of $q_t$ have been studied extensively in the Markovian case~\cite{Appert2008,Lecomte2012}.  For non-Markovian models, similar quantities have been studied in~\cite{Harris2015,Cavallaro2016}.  Given the connections between exclusion processes and traffic modelling~\cite{Nagel1996}, the generalisation of such models to include memory of previous flow (current) is quite natural~\cite{Harris2015}.

We introduce here a memory of mean-field type, so that every particle hops either right ($+$) or left ($-$) with rate $w_\pm=[1\pm \tanh(\nu q_t)]/2$, as long as the destination site is empty.  
The non-linearity in this model is similar to that of the cluster growth model which leads to some similar phenomenology.

It is useful to 
note that detailed balance is broken in this model (except for $q_t=0$),  but the dynamical rules for any given $q_t$ correspond to an asymmetric simple exclusion process with periodic boundaries, whose stationary state has all particles distributed independently (subject to the exclusion constraint).
Assuming that the system is in such a stationary state at time $t$, and its
time-averaged current is $q_t$, the (average) rate for accepted particle hops is 
\beq
 \left\langle L\frac{\rm d}{{\rm d} t} (t  q_t)  \right\rangle_{q_t} =    N \frac{L-N}{L-1} \tanh(\nu q_t) \; .
\label{equ:esep-tanh}
\eeq
Here the factor of $(L-N)/(L-1)$ is the probability that a site adjacent to a given particle is vacant.  Expanding the tanh about $q_t=0$ shows that the zero-current state $\langle q_t\rangle=0$ is stable only if $\nu <  \nu_c$ with 
\beq
\nu_c=\frac{L(L-1)}{ N(L-N) } \;.
\eeq
We identify $\nu_c$ as a {phase-transition} point, directly analogous to the cluster-growth model.
 
 \begin{figure}
\includegraphics[width=84mm]{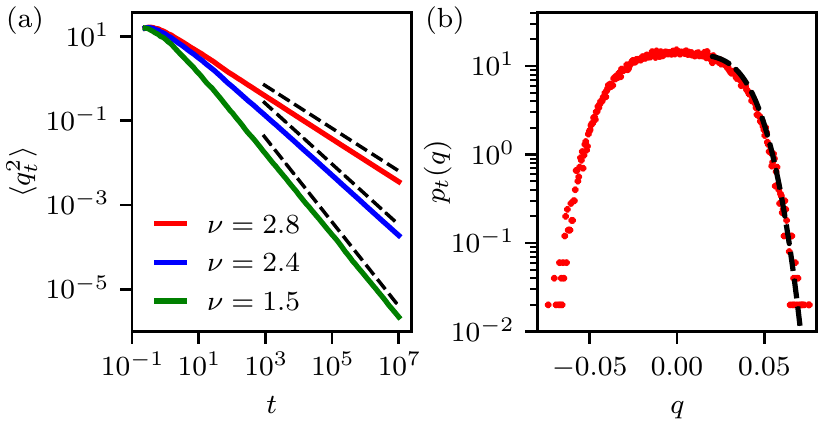}
\caption{%
Numerical results for a non-Markovian symmetric simple exclusion process with $(N,L)=(8,16)$, so $\nu_c= 3.75$.  (a)~For $\nu=2.4,2.8$, particle motion is superdiffusive so the variance of $q_t$ decays as a power law consistent with (\ref{equ:var-a}), dashed lines indicate power-law behaviour with the theoretically-predicted exponent $a=\nu/\nu_c$.  For $\nu=1.5$ the behaviour is diffusive, $\langle q_t^2\rangle \propto t^{-1}$, since $\nu<\nu_c/2$.
(b)~The distribution of $q_t$ for $\nu=2.8$ at $t=10^5$ is similar to the ERW in Fig.~\ref{fig:cgf-si}, the dashed black line is a fit to (\ref{equ:ldp-erw-smallq}) with  $a=\nu/\nu_c$. }
\label{fig:esep}
\end{figure}

For $\nu<\nu_c$, expansion of (\ref{equ:esep-tanh}) about $q_t=0$ yields (\ref{equ:vq-a})  with $a=\nu/\nu_c$, which is again similar to the 
growth model and indicates that the LIE scenario is applicable, at least for small $q^*$.  As a controlled model, we consider a (Markovian) asymmetric simple exclusion process with a time-dependent asymmetry parameter, so hops in the $(\pm)$-direction have $w_\pm=(1\pm b_t)/2$.  This controlled model also has particles distributed independently at all times. In this case the KL divergence may  be computed similarly to (\ref{equ:KL-erw}).  This allows numerical optimisation of the controlled dynamics -- the optimal behaviour is similar to the ERW and cluster growth models, showing an LIE mechanism.  An explicit LIE bound may also be derived by following exactly the same steps as used for the cluster growth model in Sec.~\ref{sec:growth}.

Contrary to the other models considered here, we do not expect this controlled model to fully  capture the large-deviation mechanism, because it neglects interparticle correlations which  are important for large deviations in exclusion processes~\cite{Appert2008}.  This effect might be captured  by combining the temporal additivity principle~\cite{Harris2015} with results for large deviations in Markovian exclusion processes~\cite{Appert2008}, but such an analysis is beyond the scope of the present work.  However, we expect the general features to be robust: a large excursion at early times and a rate function scaling as (\ref{equ:rho-LIE}).
Numerical results confirming the similarity between this non-Markovian SEP and other LIE models are shown in Fig.~\ref{fig:esep}.  
This analysis illustrates that the generic fluctuation mechanisms described in this paper are not limited to simple one-particle systems.

As a final observation,  note that since particles do not pass each other in exclusion processes, trajectories with time-averaged current $q_t=c$ at long times must have single-particle currents whose time averages  all converge to $c$ also.  For this reason, we would expect similar behaviour if each particle had an  individual memory of its own individual displacements, in contrast to the simple (mean-field) case considered here, where the motion of each particle is affected by the memory of the whole system.

\section{Outlook} 
\label{sec:outlook}

We have presented two mechanisms by which large deviations can occur in non-Markovian processes, leading to generic bounds (\ref{equ:rho-IGL},\ref{equ:rho-LIE}) on the probabilities of these rare events.  
To prove that these bounds give the right scaling in specific cases requires more detailed analysis, as illustrated here for the simple ERW and GERW models.  (Such analyses are necessary to rule out competing mechanisms with larger probability then the IGL and LIE.)
Our results indicate that the LIE mechanism operates in a non-Markovian exclusion process, and the general mechanistic insights have enabled us to clarify and extend several other results from the literature~\cite{Jack2019-growth,Harris2009,Harris2015}.  
This understanding is also relevant in socioeconomic decision models that can be approximated by generalised urn/elephant models~\cite{Harris2015-peak-end}; by revealing fluctuation mechanisms in these systems, our analysis may be utilised to predict and control their long-term fluctuations.
We look forward to future work exploiting these new insights, in order to elucidate the rich fluctuation behaviour of non-Markovian systems.

\begin{acknowledgments}
We thank Simone Franchini for helpful discussions. R.J.H. gratefully acknowledges an External Fellowship from the London Mathematical Laboratory.
\end{acknowledgments}

\begin{appendix}

\vspace{2cm}

\section{Typical fluctuations in the GERW}
\label{app:ele-prelim}

We derive (\ref{equ:var-a}) for the GERW.  
Suppose that after $t$ steps $Q_t=tq_t$ has a Gaussian distribution with mean zero and variance $v_t$.  Then $Q_{t+1}-Q_t$ is normally distributed with mean $aq_t$ and variance $1$ so the distribution of $Q_{t+1}$ is
\beq
p(Q_{t+1}) = \frac{1}{z_t} \int \exp\left[-\frac{[Q_{t+1}-(1+\frac{a}{t})Q_t]^2}{2} - \frac{Q_t^2}{2v_t} \right] \mathrm{d}Q_t
\eeq
with $z_t=\sqrt{4\pi^2 v_t}$.
This distribution is normal with mean zero and variance $v_{t+1}=1 + v_t(1+a/t)^2$.  From this recursion relation one finds a series solution for $v_t$ in terms of the gamma function:
  \begin{equation}
    v_t=\frac{\Gamma(a+t)^2}{\Gamma(a+1)^2 \Gamma(t)^2} \left( \sum_{n=1}^{t-1} \frac{\Gamma(a+1)^2  \Gamma(n+1)^2}{\Gamma(a+n+1)^2}+1 \right) \; .
  \end{equation}
The form of the large-$t$ behaviour can be obtained directly from the recursion by writing $v_t=v(t)$ so that $v'(t)\approx1+2av(t)/t$.  
Hence $v(t)\approx t/(1-2a) + ct^{2a}$ and so the variance of $q_t$ is 
\beq
{\rm Var}(q_t) = \frac{v(t)}{t^2} \approx \frac{1}{t(1-2a)} + c_a t^{-2(1-a)}
\eeq
where subleading terms at higher order in $t^{-1}$ have been omitted.  The second term is dominant for $a>1/2$ and the constant $c_a$ corresponds to $\chi_{\rm G}$ in the asymptotic variance; its value can be extracted as a limit from the series solution.  For $a=0.7$ as used in Fig.~\ref{fig:cgf-si} numerical evaluation of the limit yields $\chi_{\rm G}\approx 3.4$.

\section{Large deviations in ERW by mapping to urn model}
\label{app:urn}

For large deviations of $q_t$ in the ERW, the SCGF $\psi(\lambda)$ can be obtained exactly by adapting results of~\cite{Franchini2017}.  We state the equations and characterise the behaviour at small $\lambda$.

The ERW can be interpreted as an urn model~\cite{Baur2016}.  If the fraction of $+$ steps before time $t$ is $s_t$ then the probability that $\Delta x_{t+1}=+1$ is $\pi(s_t)$ where
\beq
\pi(s) = \frac{ 1 + a(2s-1) }{2}
\eeq
is the corresponding urn function~\cite{Franchini2017}.
Given this urn function, the parameters $(a,b)$ of Corollary 12 of~\cite{Franchini2017} correspond to $((1-a)/2,a)$ in the notation of this work.  Since $q_t=2s_t-1$ then
\beq
G(\lambda,t) = \log \langle \ee^{\lambda t (2s_t-1) } \rangle  \; .
\label{equ:G-s}
\eeq
Define $ \tilde\psi(\mu)={\lim_{t \to \infty} t^{-1} \log} \langle \ee^{\mu t s_t}\rangle$ as the SCGF of~\cite{Franchini2017}, denoted in that work by $\psi$.
Then (\ref{equ:def-psi},\ref{equ:G-s}) yield
\beq
\psi(\lambda) = \tilde\psi(2\lambda) -\lambda \; .
\eeq
Hence by Corollary 12 of~\cite{Franchini2017}  one has for $\lambda>0$ that
\beq
\psi(\lambda) = -\log\left[ 1 - w\ee^{-2w\lambda} y^{1/a} {\cal B}\left(w,-2w,y\right) \right]  -\lambda  \; ,
\label{equ:psi-BB}
\eeq
where we introduced shorthand notation $w=\frac{1-a}{2a}$ and $y=1-\ee^{-2\lambda}$ (used only within this Appendix), and where
\beq
{\cal B}(w,v,y) = \int_y^1 (1-t)^{w-1} t^{v-1} {\rm d}t
\label{equ:BB}
\eeq
is a particular case of the incomplete Beta function.  

We now compute the behaviour of $\psi$ at small $\lambda$, observing that $y\simeq 2\lambda$ in this limit. 
Our regime of interest is $1/2<a<1$ so that $0<w<1/2$.
In this case ${\cal B}(w,-2w,y)$ diverges as $y\to0$.  To extract the nature of this divergence, introduce a factor of $1=t+(1-t)$ into the integrand of (\ref{equ:BB}) to yield
\begin{align}
{\cal B}(w,v,y) & = \int_y^1 \left[ (1-t)^{w-1} t^{v} + (1-t)^{w} t^{v-1} \right] {\rm d}t
\nonumber \\
& = \frac{v+w}{v} \int_y^1 (1-t)^{w-1} t^{v} {\rm d}t - \frac{y^{v} (1-y)^w}{v} \; , 
\end{align}
where the second line used an integration by parts, with the assumption that $w>0$.  There is no such assumption on $v$, the case of interest is $-1<v<0$.  The limiting behaviour at small $y$ can now be extracted:  for $v>-1$ and $y\to0$ then
\beq
{\cal B}(w,v,y) \simeq \frac{v+w}{v} B(w,1+v) - \frac{y^{v}}{v} + o(1)  \; .
\eeq
Here $B(x,y)$ is the (complete) Beta function which is given for $x,y>0$ by $\int_0^1 t^{x-1} (1-t)^{y-1} {\rm d}t$, it is finite and positive.  Moreover, the relation
$B(w,v) =\frac{v+w}{v} B(w,1+v)$ extends the Beta function to negative arguments.
Using these results with (\ref{equ:psi-BB}) and identifying $-2w=(a-1)/a$ gives
\begin{multline}
 \psi(\lambda) = \\
 - \log\left[ 1 - w y^{1/a} 
  \left( B(w,-2w) + \frac{ y^{(a-1)/a } }{2w}  + o(1) \right) \right] 
 \\ -\lambda \; .
 \label{equ:psi-small}
\end{multline}
Finally, noting that $y\simeq 2\lambda$ and using that $\psi$ is an even function:
\beq
\psi(\lambda)  = \frac{1-a}{2a}   |2\lambda|^{1/a}  B\left(\frac{1-a}{2a},\frac{a-1}{a} \right)\left[ 1 + o(1) \right] \; .
\eeq
Hence 
\beq
c_{\rm E}= \frac{2^{(1/a)-1}(1-a)}{a}  B\left(\frac{1-a}{2a},\frac{a-1}{a} \right) 
\label{equ:cE}
\eeq
in (\ref{equ:g-erw},\ref{equ:psi-erw}).
Analysing the subleading term shows that in fact the first correction to (\ref{equ:psi-small}) is $\psi(\lambda) = c_{\rm E}|\lambda|^{1/a} + O(\lambda^2)$.

Recall that we assumed here $1/2<a<1$, since this is the regime of interest for this work.  However (\ref{equ:psi-BB}) also applies for $0<a<1/2$ -- similar analysis can also be carried out in that case.  For $a<0$ the corresponding result is given in~\cite{Franchini2017}, the resulting $\psi$ is analytic.

\section{Controlled dynamics}

\subsection{Outline of general theory}
\label{app:con-gen}

As discussed in the main text, one method for analysing fluctuation mechanisms is to construct controlled processes whose typical trajectories reproduce the rare-event behaviour of interest.  Such processes can be analysed variationally. 

We work in the generic framework where the configuration of the system at time $t$ is $\CC_t$.
A trajectory or sample path is denoted $\bm{\CC}$ and its
probability  $\bm{\CC}$ in the original model is denoted by $P(\bm{\CC})$. 
{Throughout our analysis, we fix $t$ as the trajectory length and we use $\tau$ to indicate a generic time within the trajectory.}
Now let $P_{\rm con}(\bm{\CC})$ be the probability of $\bm{\CC}$ in  some controlled model, which has different dynamics.
Optimal-control theory provides the following general inequality~\cite{Dupuis-book}
\beq
G(\lambda,t) \geq \lambda t \langle q_t \rangle_{\rm con} - {\cal D}( P_{\rm con} || P )
\label{equ:G-sup}
\eeq
where $\langle \cdot \rangle_{\rm con}$ indicates an average in the controlled model, and ${\cal D}(Q||P)$ is the Kullback-Leibler (KL) divergence between the distributions $Q$ and $P$.  To prove (\ref{equ:G-sup}) define 
\beq
P_{\rm cano}(\bm{\CC}) = \ee^{\lambda t q_t-G(\lambda,t)} P(\bm{\CC})
\label{equ:def-cano}
\eeq
 which is a normalised probability distribution, by definition of $G$.  (The subscript ``cano'' indicates that this definition is analogous to that of the canonical ensemble in thermodynamics.)  Then by definition of the KL divergence, the right-hand side of (\ref{equ:G-sup}) can be expressed as
\beq
 \lambda t \langle q_t \rangle_{\rm con} - {\cal D}( P_{\rm con} || P ) = G(\lambda,t) - {\cal D}( P_{\rm con} || P_{\rm cano} ) \; .
 \label{equ:G-eq}
\eeq
The KL divergence is non-negative so the right-hand side is less than or equal to $G(\lambda,t)$, and (\ref{equ:G-sup}) follows.  Moreover, there is equality in (\ref{equ:G-sup}) if and only if $P_{\rm con}=P_{\rm cano}$.

In addition, setting $\theta=1$ in the definition~(\ref{equ:def-psi})
we obtain $\psi(\lambda)=\lim_{t\to\infty} t^{-1} G(\lambda,t)$ so (\ref{equ:G-sup}) yields
\beq
\psi(\lambda) \geq \lim_{t\to\infty} \left[ \lambda \langle q_t \rangle_{\rm con} - \frac1t {\cal D}( P_{\rm con} || P_{\rm cano} ) \right] \; .
\label{equ:psi-sup}
\eeq
  If this bound is saturated then the controlled process gives an accurate representation of the rare event of interest, see also below.
  We emphasise that for non-Markovian processes as considered here, the limit in (\ref{equ:psi-sup}) involves controlled processes where the dynamical rule at time $\tau$ depends both on $\tau$ and on the total trajectory length $t$; accurate bounds require controlled processes with time-dependent rates.

\subsection{GERW}
\label{app:control-gerw}

We construct the optimally-controlled process for large deviations of $q_t$ in the GERW.   
Using  (\ref{equ:def-cano})  
one obtains a distribution for the trajectory $\bm{X}$, as defined in Sec.~\ref{sec:gerw-igl}:
 \beq
 P_{\rm cano}(\bm{X}) \propto \exp\left[ \lambda t q_t-G(\lambda,t)-\frac{{\cal S}(\bm{X})}{2} \right]  \; .
 \label{equ:P-cano-S}
 \eeq
where $q_t$ also depends on $\bm{X}$ though (\ref{equ:q-ele}).
This distribution is Gaussian for the increments and for the $q_\tau$, and one has an analogue of (\ref{equ:qk-micro}) which is
\beq
\langle q_\tau \rangle_{\rm cano} = \mu_\tau \langle q_t\rangle_{\rm cano} h 
\label{equ:qk-cano}
\eeq
where $h,\mu_\tau$ are the same quantities that appear in (\ref{equ:qk-micro}).   
That is, choosing $\lambda$ in the canonical ensemble fixes $ \langle q_t\rangle_{\rm cano} $.  Then the average path in this ensemble coincides with the average path in a corresponding microcanonical ensemble with $q_t = \langle q_t\rangle_{\rm cano} $.

{Since $P_{\rm cano}$ in (\ref{equ:P-cano-S}) is Gaussian, it is possible to construct exactly an optimally-controlled process that generates trajectories according to this distribution.  This process achieves equality in (\ref{equ:G-sup}) and captures the mechanism by which large rare fluctuations occur in the GERW.  This is similar to the Doob transform, as discussed in~\cite{Chetrite2015}, with time-dependent rates as in~\cite{Jack2019-growth}.}
 Within the controlled system, the displacement on step $\tau$ is Gaussian with mean $a q_{\tau-1} + b_\tau$ and variance unity.   This means that $P_{\rm con}(\bm{X})=\exp(-\tilde{S}(\bm{X})/2)$ with 
\beq
\tilde{\cal S}(\bm{X})=\sum_{\tau=0}^{t-1} [ (\tau+1)q_{\tau+1} - q_\tau (\tau+a) - b_{\tau+1} ]^2 \; ,
\eeq
analogous to (\ref{equ:S-gauss-v1}).
Hence
\beq
\tilde{\cal S}(\bm{X})= {\cal S}(\bm{X}) - 2  t  q_t b_{t} + 2 \sum_{\tau=1}^{t-1}  q_\tau [ (\tau+a) b_{\tau+1}  -  \tau b_\tau  ]  + \sum_{\tau=1}^t b_\tau^2 
\; .
\eeq
{The optimally-controlled process has $P_{\rm con} = P_{\rm cano}$ [recall (\ref{equ:G-eq})], which is achieved by} 
setting $b_{t}=\lambda$ and using $b_{\tau-1}=b_\tau(1+\frac{a}{\tau-1})$ iteratively to fix the $b_\tau$.  For the CGF this identification yields $G(\lambda,t) =  \frac12 \sum_{\tau=1}^t b_\tau^2$.

\subsection{ERW}  
\label{app:control-ele}

For the ERW, 
a variational characterisation of $\psi(\lambda)$ is available following~\cite{Franchini2017}. 
This construction also allows computation of the dominant paths shown in Fig.~\ref{fig:inst}.

 We outline the approach, which is to define a controlled process that \emph{almost} achieves equality in (\ref{equ:G-sup}), up to a correction that vanishes on taking the limit in (\ref{equ:psi-sup}). 
The {typical} path of this controlled model captures the mechanism of the (rare) fluctuations that achieve $q_t=q$ in the ERW.
(Specifically, for large $t$ and any $u>0$, the conditional distribution of $q_{ut}$ for paths that achieve $q_t=q$ is sharply peaked at $\langle q_{ut}\rangle_{\rm con}$, see~\cite{Franchini2017}.)

We use (\ref{equ:G-sup})
with the controlled dynamics described in the main text for which $(b_1,b_2,\dots,b_{t})$ are variational parameters.  The KL divergence between $P_{\rm con}$ and $P$ is
\begin{multline}
{\cal D} = \frac12 \sum_{\tau=1}^t \left[ (1+b_\tau) \log (1+b_\tau) + (1-b_\tau) \log(1-b_\tau) \right]
\\ -
\frac12 \sum_{\tau=1}^t  (1+b_\tau) \left\langle \log( 1+ aq_{\tau-1}) \right\rangle_{\rm con}
\\ -
\frac12 \sum_{\tau=1}^t  (1-b_\tau) \left\langle \log( 1- aq_{\tau-1}) \right\rangle_{\rm con} \; ,
\label{equ:KL-erw}
\end{multline}
and we have
\beq
\langle q_\tau \rangle_{\rm con} = \frac{1}{\tau} \sum_{k=1}^\tau b_k \; .
\label{equ:mean-qk}
\eeq 
Moreover, the variance of $q_\tau$ in this controlled process is at most $1/\tau$ so
it is consistent to assume that $q_\tau$ is sharply peaked for almost all terms in the sums in (\ref{equ:KL-erw}). Hence
${\cal D}\approx \hat{\cal D}$ with 
\begin{multline}
 \hat{\cal D} = \frac12 \sum_\tau \left[ (1+b_\tau) \log (1+b_\tau) + (1-b_\tau) \log(1-b_\tau) \right]
\\ -
\frac12 \sum_\tau  (1+b_\tau)  \log( 1+ \left\langle aq_{\tau-1} \right\rangle_{\rm con} ) 
\\ -
\frac12 \sum_\tau  (1-b_\tau)\log( 1- \left\langle a q_{\tau-1} \right\rangle_{\rm con} ) \; .
\label{equ:KL-erw-approx}
\end{multline}
Using (\ref{equ:mean-qk}) this is an explicit function of the $b_\tau$ variables, so the right-hand side of (\ref{equ:G-sup}) can be maximised numerically,
which yields a numerical estimate of $G(\lambda,t)$ and hence (by considering large but finite $t$) one may estimate  $\psi(\lambda)$.  

For numerical work we use a similar method to that for the GERW: we split the sums in (\ref{equ:KL-erw-approx}) into contributions from small $\tau$ and large $\tau$ and we approximate the sum over large-$\tau$ contributions by an integral (which is also estimated numerically).  This combination of sum and integral is maximised numerically to obtain estimates of $\psi(\lambda)$ and of the corresponding (average) path (\ref{equ:mean-qk}).  This yields the results of Fig.~\ref{fig:inst}.

\section{IGL mechanism in unidirectional hopping model} 
\label{app:uni}

This Appendix establishes (\ref{equ:uni-P-bound}), which means that (\ref{equ:exc-IGL}) holds for the model of Sec.~\ref{sec:uni}, with $\beta=1$.
For this condition, it is sufficient to consider a finite-time interval between $t_0$ and $\tau^*$ (there is no large-time limit because we are focussing on the excursion that occurs at early times).  For a compact notation we work on the interval $(t_0,\tau]$ and we write $k$ for a generic time within this interval.

Consider a controlled process where the first hop is at time $t_0$ (as for the original model), after which hops take place with a time-dependent rate $b(\tau)$.  Then $(\tau q_\tau-1)$ is Poissonian with mean $\int_{t_0}^\tau b(k) \mathrm{d}k$ and so
\beq
\tau\langle q_\tau\rangle_{\rm con} = 1 + \int_{t_0}^\tau b(k) \mathrm{d}k \;.
\label{equ:qb-cont}
\eeq
The KL divergence of (\ref{equ:G-sup}) is 
\beq
{\cal D} = \int_{t_0}^\tau\left\{ b(k) \left\langle \log \frac{b(k)}{a q_k} \right\rangle_{\rm con} - b(k)  + \langle a q_k \rangle_{\rm con} \right\} \mathrm{d}k \; ,
\eeq 
similar to (\ref{equ:KL-erw}).  In addition to (\ref{equ:G-sup}), the KL divergence also allows a bound on the probability distribution of $q_t$.
Roughly speaking, if one can construct a controlled process such that {the large-deviation event occurs with probability one},  $P_{\rm con}(q_\tau \geq q)=1$, 
then the probability of this event in the original model can be bounded from below:
\beq
-\log P(q_\tau \geq q) \leq {\cal D}(P_{\rm con}||P) \;.
\label{equ:P-inf}
\eeq
This may be proved by Jensen's inequality; a more precise statement is given (for example) in Equs.~(14,15) of~\cite{Jack2019-growth}.  Hence we seek an upper bound on ${\cal D}$.

To achieve this, we use $\log(1/x)\leq(1/x)-1$ with $x=q_k/\langle q_k\rangle_{\rm con}$ to write
\begin{multline}
{\cal D} \leq \int_{t_0}^\tau \Bigg\{ b(k) \log \frac{b(k)}{a  \langle q_k\rangle_{\rm con}}  -2 b(k)+ \langle a q_k \rangle_{\rm con} 
 \\
+ b(k) \langle q_k\rangle_{\rm con} \left\langle \frac{1}{q_k} \right\rangle_{\rm con} \Bigg\} \mathrm{d}k\;.
\label{equ:D-uni-v1}
\end{multline}
\newcommand{\xbar}{\overline{x}}
For a Poisson random variable $X$ with mean $\xbar$, one has 
$\langle \frac{1}{1+X}\rangle = \ee^{-\xbar} \sum_{n=0}^\infty \xbar^n / (n+1)! = (1-\ee^{-\xbar})/\xbar$.
Since $(kq_k-1)$ is Poissonian, we obtain
\begin{multline}
{\cal D} \leq \int_{t_0}^\tau \Bigg\{ b(k) \log \frac{b(k)}{\langle aq_k\rangle_{\rm con}}  - 2b(k) + \langle a q_k \rangle_{\rm con} 
 \\
 + b(k)  \langle k q_k\rangle_{\rm con} \frac{1-\ee^{-\langle kq_k-1\rangle_{\rm con}}}{\langle kq_k-1\rangle_{\rm con}}  \Bigg\} \mathrm{d}k \;.
\label{equ:D-uni-v2}
\end{multline}
{To recover the results of~\cite{Harris2015} one should assume that  $k q_k\gg 1$ throughout the integration range, so that the second line of the integrand reduces to $b(k)$.  This is valid for $t_0\gg1$. Then one sets $\tau=t$ and} minimises the resulting KL divergence over the path $\hat{q}(k) = \langle q_k \rangle_{\rm con}$, using (\ref{equ:qb-cont}) to replace $b(k)\to (\partial/\partial k) (k\hat{q}(k))$.  
The optimal path behaves for short times as $k \hat{q}(k) = 1+A[(k/t_0) - 1]$ where $A$ is proportional to the size of the giant excursion~\cite{Harris2015}.

Our approach here does not require $t_0$ to be large: we retain all terms in (\ref{equ:D-uni-v2}), and use (\ref{equ:P-inf}) with $\tau=\tau^*$
to establish (\ref{equ:exc-IGL}).
To obtain a convenient bound we set $\tau^*=2t_0$ and choose $b(k)$ such that 
$
 \langle k q_k\rangle_{\rm con} = 1 + A x 
$
  with $x=(k/t_0)-1$  and $A=2q^*t_0-1$. 
  This requires  $b(k)=A/t_0$ and ensures that $\langle q_{\tau^*}\rangle_{\rm con}=q^*$.
  [Note, $b(k)$ is only independent of $k$ for ${k}<\tau^*$ (i.e., during the excursion), the controlled process reverts to the natural dynamics of the model for ${k}>\tau^*$.]
Then (\ref{equ:D-uni-v2}) with $\tau=\tau^*$ becomes
\begin{multline}
{\cal D} \leq A \int_{0}^1 \Bigg\{ \log \frac{A(1+x)}{a  (Ax+1) }  -2 + \frac{a(Ax+1)}{A(1+x)} 
\\
+ (1+Ax) \frac{1-\ee^{-Ax}}{Ax}
 \Bigg\} \mathrm{d}x \;.
\end{multline}
We are concerned with the limit $q^*\to\infty$ which corresponds to $A\to\infty$.  {The integral can be evaluated in this limit and} the KL divergence scales as
\beq
{\cal D} \lesssim \gamma_{\rm uni} q^* t_0
\label{equ:D-leading}
\eeq
with $\gamma_{\rm uni}=2 [ \log(4/a) +a (1-\log 2) - 1]$.  
To apply (\ref{equ:P-inf})  we require additionally that $P_{\rm con}(q_{\tau^*}\geq q^*)\to1$ as $q^*\to\infty$: this holds because the distribution of $ q_{\tau^*}$ is Poissonian with a diverging mean equal to $q^*$, so it is sharply peaked at $q^*$.
Hence (\ref{equ:P-inf}) is applicable with KL divergence (\ref{equ:D-leading}) and the probability of the excursion obeys (\ref{equ:uni-P-bound}), as required.

\end{appendix}

\bibliographystyle{apsrev4-1}
\bibliography{dev}
\end{document}